\documentclass[prd,aps,twocolumn,showpacs,floats,floatfix]{revtex4}
\usepackage[dvips]{graphicx}
\usepackage{url}
\begin{document}
 
\title{Computational Relativistic Astrophysics With Adaptive Mesh Refinement: 
Testbeds}

\author{Edwin Evans${}^{(1)}$, Sai Iyer${}^{(1)}$, Erik
Schnetter${}^{(2)}$, Wai-Mo Suen${}^{(1,3)}$, Jian Tao${}^{(1)}$,
Randy Wolfmeyer${}^{(1)}$, and Hui-Min Zhang${}^{(1)}$}

\affiliation{${}^{(1)}$McDonnell Center for the Space Sciences,
Department of Physics, Washington University, St.~Louis, Missouri
63130} 
\affiliation{${}^{(2)}$Albert-Einstein-Institut, Am M\"uhlenberg
1, D-14467, Potsdam, Germany} 
\affiliation{${}^{(3)}$Physics
Department, University of Hong Kong, Hong Kong}

\date{\today}
 
\begin{abstract} 

  We have carried out numerical simulations of strongly gravitating
  systems based on the Einstein equations coupled to the relativistic
  hydrodynamic equations using adaptive mesh refinement (AMR)
  techniques.  We show AMR simulations of NS binary inspiral and
  coalescence carried out on a workstation having an accuracy
  equivalent to that of a $1025^3$ regular unigrid simulation, which
  is, to the best of our knowledge, larger than all previous
  simulations of similar NS systems on supercomputers.  We believe the
  capability opens new possibilities in general relativistic
  simulations.

\end{abstract}


\maketitle

\paragraph{Introduction} 

Numerical study of compact systems has received much attention due to
observations in high-energy astronomy and the promise of gravitational
wave astronomy.  Most effort focuses on solving the Einstein equations
with finite differencing methods.  The main difficulty of this
approach is that many general relativistic astrophysical processes of
interest, e.g., processes involving black holes and neutron stars,
require computational resources that are beyond
what present day computers allow.  The reasons that they are
computationally demanding are 1. the lack of symmetry in realistic
astrophysical situations, requiring the solving of the full set of
Einstein equations coupled to the general relativistic hydrodynamic
(GRHydro) equations in 3+1 dimensional spacetime; and 2. the
involvement of many length scales.

The difficulty of multiple length scales can be illustrated with the
neutron star (NS) coalescence problem, one of our main systems of
study.  The length scales involved are: (i) A short length scale
coming from the internal dynamics of a neutron star as a self
gravitating object. One needs to resolve the density/pressure
variation accurately enough to maintain a stable configuration in the
Einstein theory.  (ii) A longer length scale coming from the dynamics
of two NSs moving under the influence of one another, i.e., the length
scale of the orbital radius. (iii) The dynamical time scale of the
system (the orbital period $T$) turns into a long length scale due to
the dynamical nature of Einstein gravity (no such difficulty exists in
Newtonian gravity, where one can evolve orbiting system more easily).
The space surrounding the NSs within the corresponding length scale
(the wavelength of the gravitational wave due to the orbital motion)
needs to be covered in the computational domain, both for the
extraction of the waveform {\it and} for an accurate dynamical
evolution (the problem manifests itself as that the evolution can be
affected by the outer boundary if put too close).  (iv) The secular
evolution time scale of the orbital motion turns into a resolution
requirement for the numerical simulation as computational error
accumulates.  Our study in a previous paper \cite{Miller03b} indicates
that

1. To simulate a single isolated NS in a stable fashion with the
   Einstein equations requires a resolution on the order of $0.1 M_0$,
   where $ M_0 $ is the baryonic mass of the NS, for a typical
   equation of state (EOS).

2. To set initial data in a fashion (e.g., using the conformally flat
   quasi-equilibrium (CFQE) approach) that we can have some confidence
   of its being astrophysically relevant, the
   initial separation of the two NSs would have to be on the order of
   $50 M_0$ (depending on the initial spin states of the two NSs).
 
3. To get inspiral dynamics without much artificial influence
   from the boundary of the computational domain, it
   has to be put at least $0.5 \lambda $ away, where $\lambda$ is the
   gravitational wavelength of the system (assuming the present state of
   the art in setting outer boundary conditions for the constrained
   system of the Einstein equations).

4. To be able to accurately extract a gravitational waveform from the
   simulation, the computational domain should include up to $1
   \lambda$.

5. To be able to evolve the spiraling NSs within the convergence
   regime to the point of coalescence: This depends on the choice of
   initial configuration and the numerical method used.  With all
   existing methods we know of, the longer in time one needs to stay
   within the convergence regime (i.e., the constraint violations and
   physical quantities converging with respect to increasing
   resolution throughout that time period), the finer the resolution
   has to be.  In our simulation reported in \cite{Miller03b}, a
   $643^3$ simulation with $\Delta x =0.2 M_0$, covering up to $0.28
   \lambda$ for orbiting NSs at an initial separation of $28 M_0$
   (with an angular frequency of $\Omega =0.012 {M_0}^{-1}$), the
   system remains in the convergence regime for only about half an
   orbit.  Being so far off from our target of evolving to the
   coalescence point, an estimate of what might be needed would be
   meaningless.

The wavelength of a gravitational wave with orbital separation of $50
M_0$ (cf., (2) above) is about $1,000 M_0$.  For a unigrid at $0.2
M_0$ (cf., (1) above), the requirements (3) and (4) imply a grid of
$5,000^3$.  A computer with a memory size capable of doing such a
simulation will not be available in the near future.

Hence the biggest obstacle we encounter in NS coalescence simulation
based on finite differencing of the Einstein equations is the need for
a large number of grid points, which translates into large computer
memory and long execution time.  We need the adaptive mesh refinement
(AMR) treatment: Use fine grid patches co-moving with the compact
objects to satisfy the resolution required by (1), and a coarse grid
extending to the local wave zone for (2),(3) and (4).  Similar
considerations have motivated much effort in this direction, see e.g.,
\cite{amr_others} for recent progress.

Unfortunately, application of AMR techniques in general relativistic
astrophysics is more difficult than one might naively think.  Although
the theory and algorithms of mesh refinement are well established in
computational science, and the numerical treatments of the Einstein
and GRHydro equations have been extensively investigated by
relativists and astrophysicists, after many years of intense effort by
many research groups it has not been possible to put the two together
for a fully general relativistic 3D AMR simulation.  The main
difficulty is that it involves huge infrastructures on {\it both} the
computer science side and the physics side: it is difficult for
computer scientists to dive into the complexity of the physics, and
vice versa.  As a rough representation of the complexity, in our code
construction process, we have to integrate a 100,000 line mesh
refinement code (GrACE \cite{GrACE}), a 85,000 line general
relativistic astrophysics code (GR-Astro \cite{GRAstro}) and a 500,000
line parallel computational library (Cactus Toolkit \cite{Cactus})
that GR-Astro makes use of.  One central message of this paper is: We
confirm that there is no issue of principle involved in enabling
general relativistic AMR, the devils are all in the details.

In this paper we demonstrate for the first time that a full 3+1
dimensional simulation based on the Einstein theory can
be carried out with AMR.  Three sample systems are studied:

1. A NS moving at a speed of $0.5c$ described by the Einstein plus
GRHydro equations. The validity of our AMR treatment is examined with
convergence tests.  Convergence tests are more complicated with AMR;
three different kinds of convergence tests are presented: (i)
simulations with increasing resolutions on all grid levels, (ii)
simulations with added levels of refinement, and (iii) comparison to
unigrid results.  The investigation of a boosted star, which invokes
{\it all} terms in the evolution equations, played an important role
in our code construction process.

2. Two NSs coalescing with angular momentum ($L = 5.9 {M_\odot}^2$).
The study demonstrates that our AMR treatment can (i) handle
collisions and merging of not only NSs, but also grid patches, (ii)
handle gravitational collapses, and (iii) simulate NS processes with
an accuracy comparable to that of a unigrid run with resolution same
as the resolution of the finest grid of the AMR run.

3.  An inspiraling NS binary. The two NSs are covered by co-moving
fine grid patches, with the coarsest grid covering a fraction of a
wavelength of the system.  We show an AMR simulation which is
equivalent to a regular $1025^3$ unigrid simulation, larger than any
simulation of NS binary systems performed so far.

In the following sections we discuss these 3 simulations.  The last
section summarizes and discusses the next steps. 

\paragraph{Boosted Neutron Star.} 

We begin with a study of a NS moving across an otherwise empty space
at a constant speed.  Although the physical system is not changing in
time beyond a uniform boost, the metric has complicated spacetime
dependences due to the frame dragging effect.  Accordingly, all
coordinate quantities including those of the spacetime and matter are
changing in a non-trivial manner (not just a uniform translation). In
the simulation, we start with a configuration satisfying the
Hamiltonian constraint (HC) and momentum constraint (MC) representing
a NS boosted to $0.5c$, and evolve it with the full set of dynamical
Einstein equations coupled to the GRHydro equations.  The system of
equations as well as the conventions we use in this paper are given in
\cite{Miller03b}.  The simulation provides a good test for our code as
it invokes {\it all} terms in the equations, and is numerically a
fully dynamical test.
 
The NS is described by a polytropic EOS: $P = (\Gamma - 1) \rho
\epsilon $ with $\Gamma = 2$ ($P = k \rho^\Gamma$ for initial data,
with $k = 0.0445 c^2/\rho_n$, where $\rho_n$ is the nuclear density,
approximately $2.3$ x $10^{14} \; \rm{g/cm}^3$).  (All simulations
reported in this paper use the same EOS.) The NS has a proper radius
of $R=12 M_\odot$, an ADM mass of $1.4 M_\odot$ and a baryonic mass $
M_0=\frac {1}{2} \int d^3\!x \, \sqrt{\gamma} \rho W =1.49 M_\odot$.
(For these values of the parameters, the maximum stable NS
configuration has an ADM mass of $1.79 M_{\odot}$ and a baryonic mass
of $ 1.97 M_{\odot}$).  The initial data is obtained by imposing a
boost on the TOV solution (\cite{Font00}).  The evolution is carried
out with the $\Gamma$ freezing shift and the ``$1+\log$" lapse (for
details of the shift and lapse conditions and method of
implementations, see \cite{Miller03b}).


\begin{figure*}
\hfill
\begin{minipage}[t]{.3\textwidth}
\begin{center}
\includegraphics[width=2in]{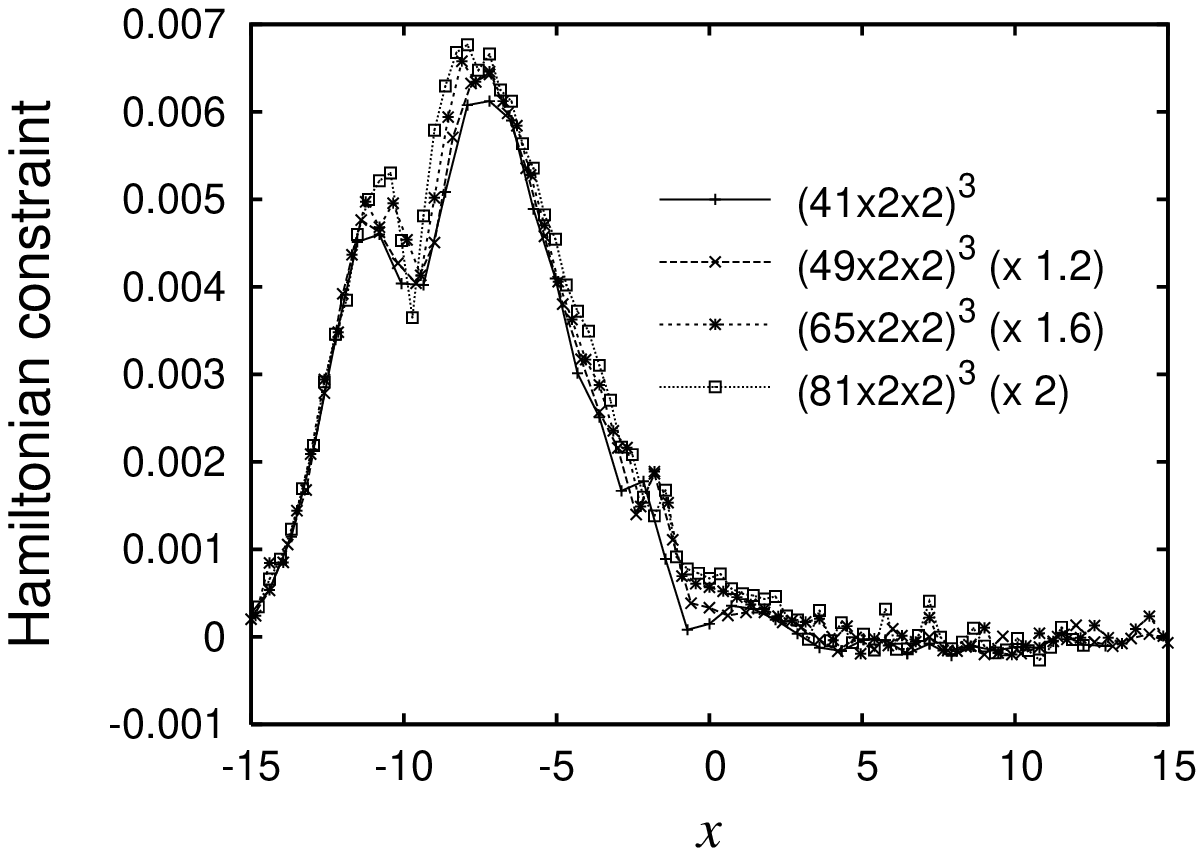}  
\vspace{-0.2in}
\caption{HC violation scaled by $(1/dx)$ for runs with different
  resolutions. Their overlapping with one another implies first order
  convergence (as the TVD hydro scheme dictates).}
\label{fig:boost_ham_conv}
\end{center}
\end{minipage}
\hfill
\begin{minipage}[t]{.3\textwidth}
\begin{center}  
\includegraphics[width=2in]{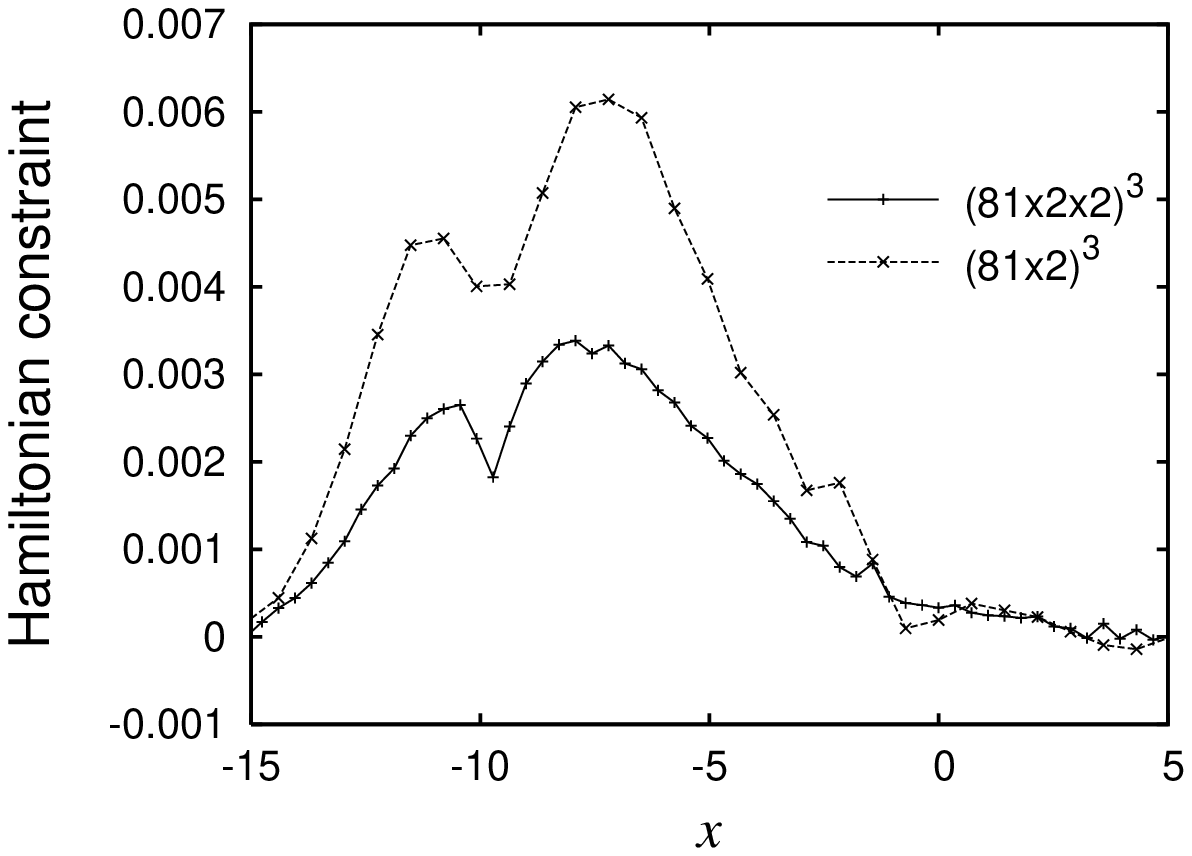}
\vspace{-0.2in}
\caption{Effect of more levels: the addition of one refinement level
  lowers the HC violation by a factor of 2 in the region of extra grid
  level where the HC violation is significant.}
\label{fig:boost_ham_lev}
\end{center}
\end{minipage}
\hfill
\begin{minipage}[t]{.3\textwidth}
\begin{center}  
\includegraphics[width=2in]{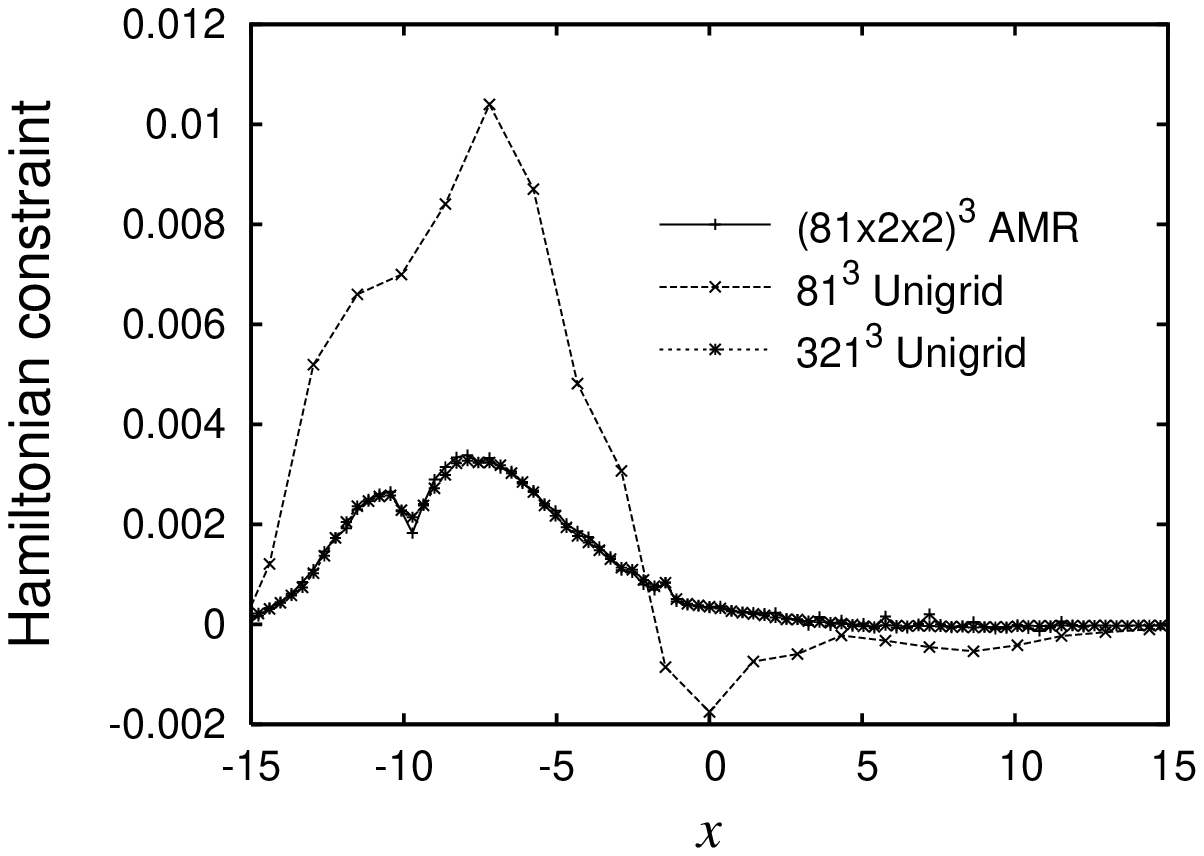}
\vspace{-0.2in}
\caption{The HC violation of AMR run coincides with that of the
  unigrid $321^3$ run which has a resolution the same as the finest
  grid of the AMR run. }
\label{fig:boost_ham_uni}
\end{center}
\end{minipage}
\hfill
\vspace{-0.25in}
\end{figure*}

The computational grid is set up as follows: 1.~The coarse grid has a
resolution of $dx=2.88M_\odot$ (4 points across the radius $R$ of the
NS) covering a region of $(58M_\odot)^3$.  2.~Two levels of adaptive
fine grid with $dx=1.44 M_\odot$ and $dx=0.72 M_\odot$ are set up.
The adaptive grid is allowed to change in size and location as the
refinement criteria dictate.  3.~Two different refinement criteria
have been studied: (i)~value of matter density $\rho$, and (ii)~amount
of HC violation.  Combinations of the two with ``or" can be used.  It
turns out that for the neutron star studies it does not matter much
which condition is used: the central region of the NS is at the same
time the region of highest density, maximum HC violation and maximum
evolution error.  All simulations shown in this paper are obtained
with (i).


Fig.~\ref{fig:boost_ham_conv}--\ref{fig:boost_ham_uni} examine the
validity of the simulation with 3 kinds of convergence tests.
Fig.~\ref{fig:boost_ham_conv} shows the violation of the HC at
$t=28.8M_\odot$ along the $x$-axis for four different runs.  The HC
violation is calculated on the finest grid available for regions
covered by more than one grid (as for all HC plots in this paper).
The resolutions of the runs are $41^3$, $49^3$, $65^3$, and $81^3$,
corresponding to $dx=2.88M_\odot, 2.4M_\odot, 1.8M_\odot, {\rm and }
1.44M_\odot $, respectively, on the base grid.  (The notation
$(41\times2\times2)^3$ indicates a $41^3$ base grid and two levels of
refinement with a refinement ratio of 2 each.) The results for the
higher resolution runs have been scaled linearly.  The plot
demonstrates that the code is converging to first order, which is the
expected rate of convergence as we used a high resolution shock
capturing TVD scheme \cite{Miller03b} in our hydrodynamic evolution
which is first order at extremal points.

In fig.~\ref{fig:boost_ham_lev}, we compare the HC violations of two
runs at time $t=28.8M_\odot$: (i)~the $(81\times2\times2)^3$ run shown
in fig.~\ref{fig:boost_ham_conv}, and (ii)~an $(81\times2)^3$ run with
only one level of refinement covering the high density region.  We see
that the addition of a refinement level lowers the HC violation by a
factor of 2 in the region of the extra grid level (where the HC
violation is significant).


\begin{figure*}
\includegraphics[width=2.25in,clip=true, bb= 90 165 365 290]
  {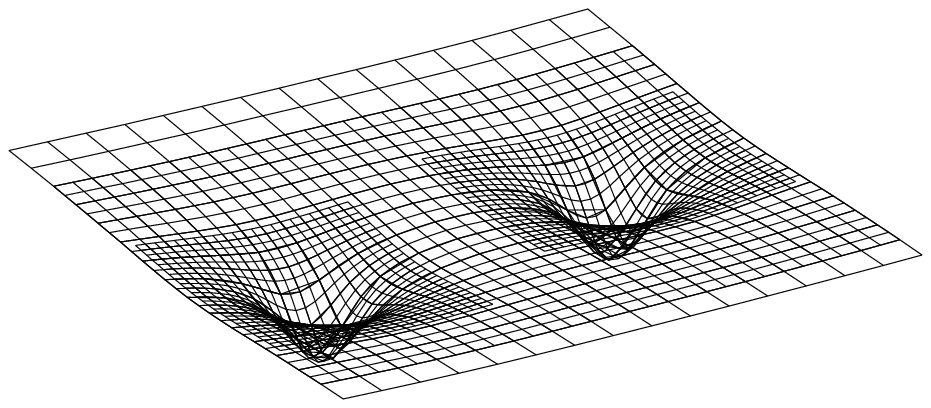}%
\includegraphics[width=2.25in,clip=true, bb= 90 165 365 285]
  {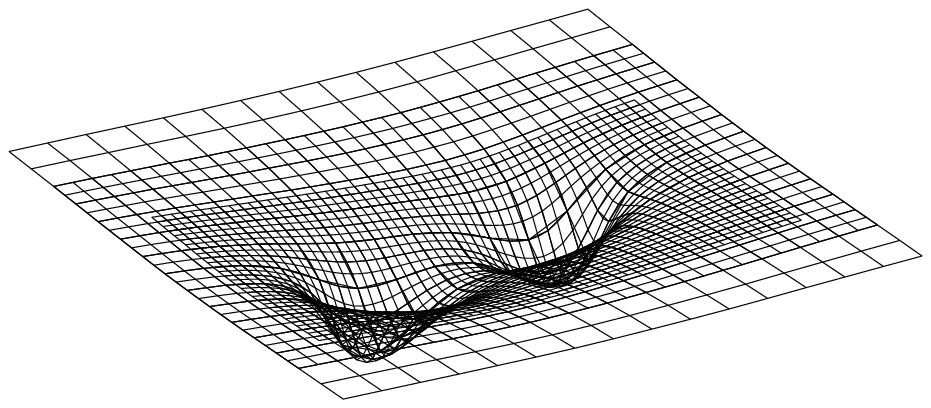}%
\hspace{-0.35in}
\includegraphics[width=2.25in,clip=true, bb= 90 135 365 290]
  {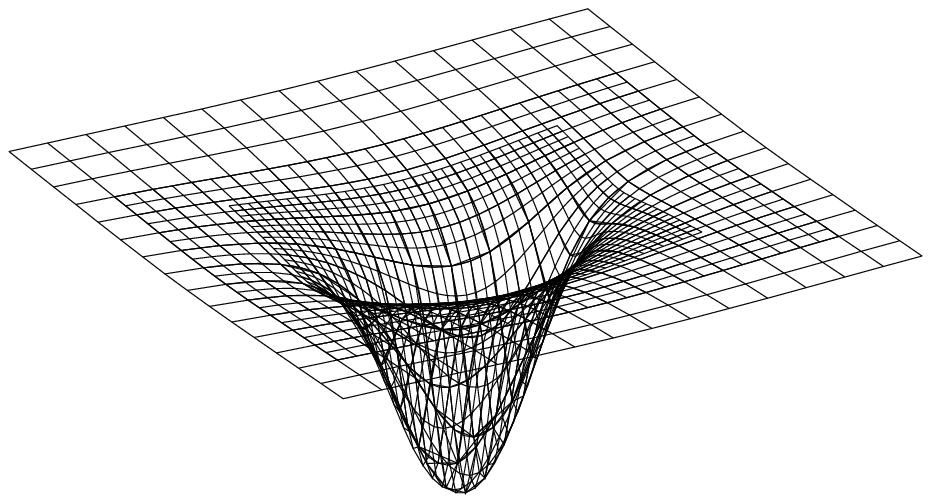}
\vspace{-0.15in}
\begin{center}
(a) $t=0$\hspace{1.5in}(b) $t=61.2{M_\odot}$\hspace{1.5in}(c) $t=122.4{M_\odot}$
\end{center}
\caption{\label{fig:offaxis_alp} The lapse of coalescing NSs at 3
  different times, with the grid structure superimposed, and with 1 to
  2 downsampling (showing every other point).  Only the inner part of
  the computational domain is shown.}
\end{figure*}

In fig.~\ref{fig:boost_ham_uni}, we show the HC violations of three
runs at $t=28.8M_\odot$.  The AMR run is again the
$(81\times2\times2)^3$ one in fig.~\ref{fig:boost_ham_conv}. The other
two are unigrid runs, one ($81^3$) at the resolution of the coarsest
AMR grid ($dx=1.44M_\odot$), and the other ($321^3$) at the resolution
of the finest AMR grid ($dx=0.36M_\odot$).  We see that the AMR run
has exactly the same accuracy as the unigrid fine resolution run (the
two lines coincide).  This is an important point for our study: For NS
simulations in this and the following sections, finite difference
error is most significant in the high density region covered by the
finest AMR grid; using coarser grids elsewhere does not affect the
accuracy of the simulation.  The fact that the error can be the same
for an unigrid run and an AMR run with suitable fine grid coverage
enables us to speak of the ``unigrid equivalent" of an AMR run: a
unigrid run with the resolution of the finest AMR grid.

The three kinds of convergence tests provide confidence in the
validity of our AMR treatment.

\paragraph{Coalescing Neutron Stars.} 

In this section we study the coalescence of two NS's having baryonic
masses and EOS as in the boosted star case above.  The NSs have their
equatorial plane on the $x$-$y$ plane and an initial center to center
(points of maximum mass) separation of $5R$ in the $x$ and $0.83R$ in
the $y$ directions ($R=12 M_\odot$).  The NSs are boosted in the $+/-
x$ directions with the total angular momentum of the system equal to
$2.67 {M_0}^2=5.9 {M_\odot}^2$.

To determine the initial data we solve the HC and MC equations on a
unigrid of $(385, 257, 257)$ at a resolution of $dx=0.9{M_\odot}$.
The initial metric and hydrodynamic data are then interpolated onto
the AMR grids.  The AMR simulation is then compared to the unigrid
one.  The $\Gamma$ freezing shift and $1+\log$ lapse are
used in both cases.


In fig.~\ref{fig:offaxis_alp} we show the lapse (represented as height
fields) on the equatorial plane of the NSs at 3 different times $t=0,
61.2, 122.4{M_\odot}$ in the AMR simulation, with the grid structure
superimposed (downsampled by a factor of 2, and only the inner part is
shown).  We see initially there are two separated fine grid patches.
At $t=61.2{M_\odot}$, the two NSs, as well as their respective fine
patches, begin to merge.  At $t=122.4{M_\odot}$ a black hole has
formed with the lapse dipping to $0.002$ at the center, and the fine
grid patches have completely merged and shrunk into a cube.

\begin{figure}[b]
\hfill
\begin{minipage}[t]{.45\columnwidth}
\begin{center}
\includegraphics[width=1.75in]{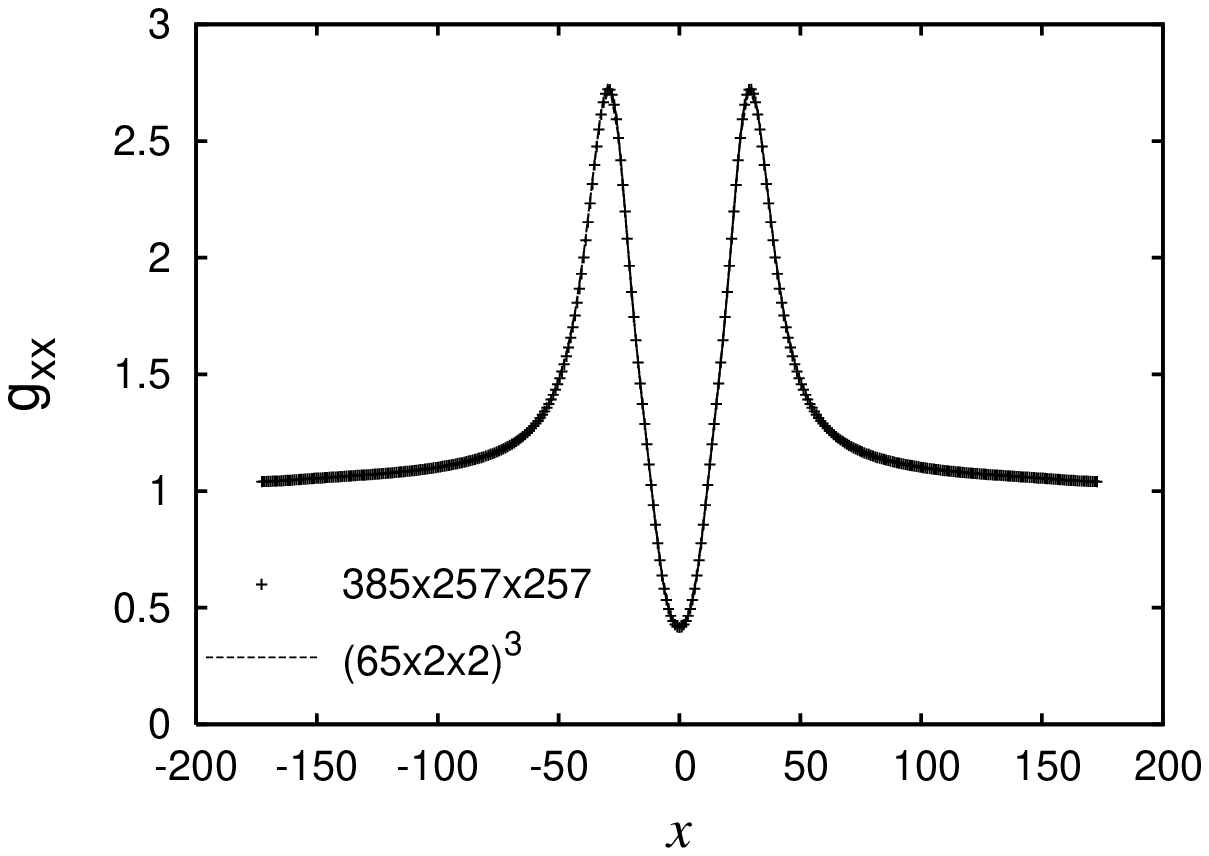}\\
(a)
\end{center}
\end{minipage}
\hfill
\begin{minipage}[t]{.45\columnwidth}
\begin{center}  
\includegraphics[width=1.75in]{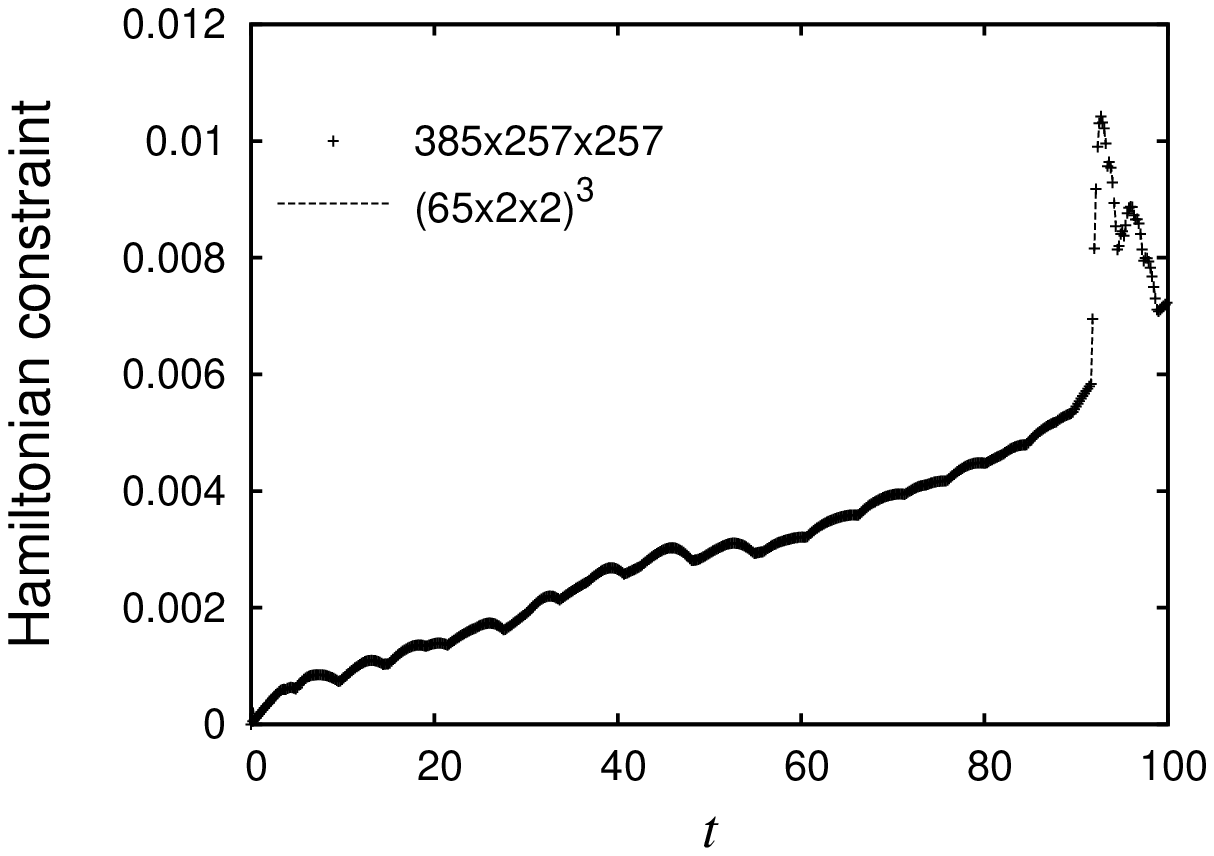}\\
(b)
\end{center}
\end{minipage}
\hfill
\caption{\label{fig:offaxis_gxx_ham}The AMR (dashed) and unigrid
  (daggers) simulations give the same results for coalescing NSs:
  (a)$g_{xx}$ at $t=122.4M_\odot$, (b) HC violations vs time.}
\end{figure}

In fig.~\ref{fig:offaxis_gxx_ham}a we plot $g_{xx}$ along the $x$ axis
at $t=122.4M_\odot$ for the AMR simulation (dashed line) and the
unigrid simulation (daggers) that has the same resolution) as the
finest AMR grid.  We see the results coincide to high accuracy.  All
metric functions and hydro variables show the same degree of agreement
even at this late time.  In fig.~\ref{fig:offaxis_gxx_ham}b we compare
the values of the (spatial) maximum of the HC violations (which is one
of the most sensitive measure of differences between runs) over time.
We see that the two simulations give basically the same results
throughout the evolution.

\paragraph{Inspiraling Neutron Stars.} 

In this section we demonstrate that with AMR we can now carry out on a
workstation (Dell Poweredge 1850) NS inspiral simulations that are
beyond existing unigrid simulations on supercomputers.

The NSs are taken to be initially in a conformally flat
quasi-equilibrium (CFQE) irrotational circular orbit with an orbital
separation of $3.3R$.  Each NS has a baryonic mass of $1.625
{M_\odot}$ with the same EOS as before.  The initial data
is obtained by solving the CFQE equations using the pseudo-spectral
code developed by the Meudon group \cite{Gourgoulhon01,Taniguchi02},
and imported onto the Cartesian grid structure in GR-Astro-AMR for
dynamical evolutions, again with $\Gamma$ freezing shift and $1+\log$
lapse.

\begin{figure}[b]
\includegraphics[width=2.5in]{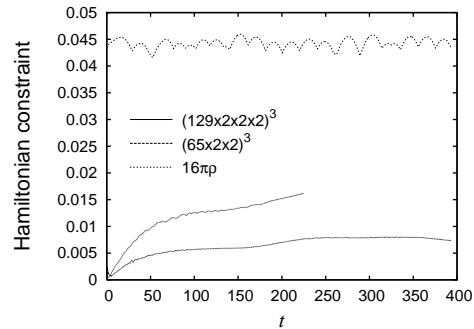}
\vspace{-0.1in}
\caption{Maximum HC violations for inspiraling NSs at two different
  resolutions, showing first order convergence. The maximum of
  $16\pi\rho$ (dotted line) is given for comparison.}
\label{fig:cfqe_ham}
\vspace{-0.2in}
\end{figure}

\begin{figure*}
\includegraphics[width=3.5in, clip=true, bb= 125 80 320 270]
  {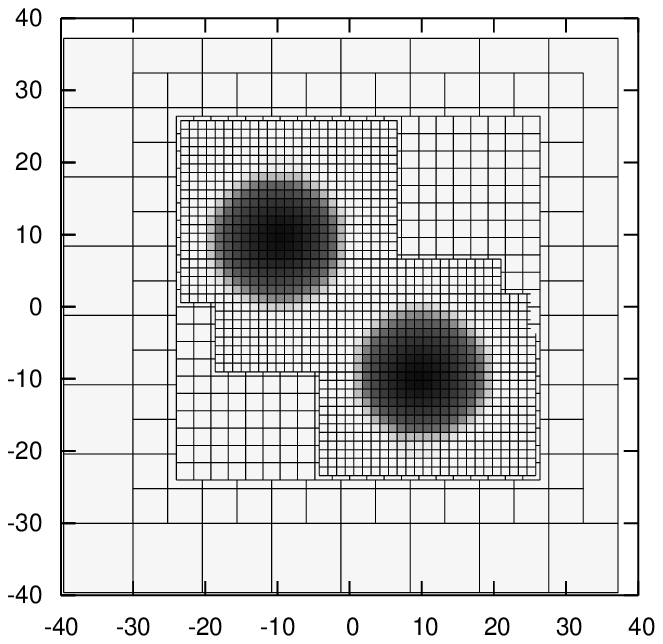}%
\includegraphics[width=3.5in, clip=true, bb= 125 80 320 270]
  {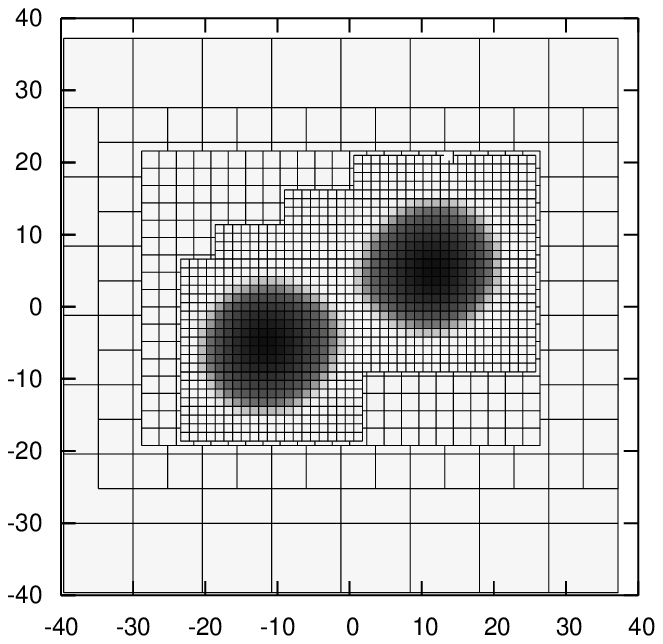}
\vspace{-0.2in}
\caption{\label{fig:cfqe_rho}Density for inspiralling NSs is given in
  grayscale with 4 levels of grid structure (downsampled by a factor
  of 4) superimposed. Only the central part of the $(34R)^3$
  computational domain is shown. (a) $t=288M_{\odot}$, (b)
  $t=418M_{\odot}$ }
\vspace{-0.2in}
\end{figure*}

We show results from two AMR simulations.  In the high resolution run
with 4 levels of refinement (with refinement ratio 2), the finest level
grid has 60 points across each NS, and the coarse grid has $129^3$
covering a computational domain of $(34R)^3$.  The low resolution run
uses 3 levels of refinement and a coarse grid of $65^3$, with the
finest level having 30 points across each NS.  The maximum HC
violations for the two runs are compared in fig.~\ref{fig:cfqe_ham},
where the maximum value of $16\pi \rho$ is also shown. ($16\pi \rho$
is the source term in the HC equation.)  We see that the error is
converging to 1st order as expected. In fig.~\ref{fig:cfqe_rho}, the
density on the equatorial plane in a gray scale plot with the grid
structure imposed is shown at $t=288 {M_\odot}$ and $t= 418{M_\odot}$
for the high resolution 4 level AMR run, with a downsampling factor of
4 (showing one in every 4 points).  We have zoomed in to show the
inner part of the computational domain.  At $t=0$ (not shown) the two
finest grid patches are separated.  By $t=288 {M_\odot}$ the
separation between the NSs has shrunk and the fine patches become
connected, but the NSs remain basically spherical.  By $t=
418{M_\odot}$, the separation between the NSs has decreased
significantly and tidal deformation is now visible.  Notice also the
change of shape of the two intermediate level grids as the NSs rotate.
The run was terminated at $t= 418{M_\odot}$ (not crashing).

\paragraph{Discussions and Conclusions.} 

We discussed the necessity of having AMR capability in general
relativistic simulations of NS coalescences.  We demonstrated that the
GR-Astro-AMR Code is capable of such AMR simulations.  The simulations
presented in this paper are the first steps towards what is needed in
NS inspiral coalescence studies.

In sec.~b we showed some of the validation studies we carried out
(based on various kinds of convergence tests) for GR-Astro-AMR with a
boosted NS.  In sec.~c we demonstrated that GR-Astro-AMR can be used
to simulate NS coalescences and formation of black holes, with an
accuracy comparable to that of an unigrid simulation using a
resolution same as that of the finest grid in the AMR run.  Sec.~d
showed an AMR simulation of an inspiraling NS binary carried out on a
Dell workstation with 8~GB of memory, which is equivalent in accuracy
to a $1025^3$ unigrid run that requires over 1.2TB of memory.  To the
best of our knowledge this is larger than all previous simulations of
similar systems on supercomputers.

In a future publication, we will extend the study in sec.~c to analyze
the amount of matter available for accretion after the NS
coalescence/BH formation, as a function of the angular momentum of the
system at the plunge point of the inspiral.  We will extend the study
in sec.~d to determine astrophysically realistic initial data for
inspiral, following the line initiated in \cite{Miller03b}.  These
investigations require more computational resources than are available
to us if they are to be carried out in unigrid.

There are many aspects in the GR-Astro-AMR code that need improvement
as a computational infrastructure for general relativistic
simulations.  In the next steps, we will (i) develop the parallel
capacity of GR-Astro-AMR, (ii) study the usage of different refinement
criteria for other NS/BH processes, and (iii) enable the direct
solving of elliptic equations on the grid hierarchy.

The code is developed with the intention of providing a computational
tool to the general relativistic astrophysics community.  The unigrid
version of GR-Astro has been released (available at
http://www.wugrav.wustl.edu). GR-Astro-AMR will be released as soon as
ready.  We invite researchers to join us in making use of as well as
further developing this code.

\acknowledgments

GR-Astro is written and supported by Mark Miller, Hui-Min Zhang, Sai
Iyer, Ed Evans, Philip Gressman and others.  The AMR version
GR-Astro-AMR and the PAGH driver it uses is written and supported by
Erik Schnetter, Ed Evans and Randy Wolfmeyer.  The version of GrACE
library originally by Manish Parashar was adapted for GR-Astro-AMR by
Ed Evans and Randy Wolfmeyer.  The Cactus Toolkit is by Tom Goodale
and the Cactus support group.  We thank Eric Gourgoulhon, Ian Hawke
and Luca Baiotti for help with the Meudon CFQE initial data.  The
research is supported in parts by NSF Grant Phy 99-79985 (KDI
Astrophysics Simulation Collaboratory Project), NSF NRAC MCA93S025,
DFG grant SFB 382 and the McDonnell Center for Space Sciences at the
Washington University.

\bibliographystyle{prsty}

\begin{thebibliography}{1}

\bibitem{Miller03b}M.~Miller and P.~Gressman and W.-M.~Suen,
Phys.\ Rev.\ D {\bf 69}, 064026 (2004)
 
\bibitem{amr_others} B.~Imbiriba et al. gr-qc/0403048; E.~Schnetter et
  al., Class.\ Quant.\ Grav.\ 21, 1465 (2004); D.~Choi et al., J.\
  Comput.\ Phys.\ 193, 398 (2004); F.~Pretorius and L.~Lehner J.\
  Comput.\ Phys.\ 198, 10 (2004); P.~Diener et al., Class.\ Quant.\
  Grav.\ 17, 435 (2000); K.~C.~B.~New, Phys.\ Rev.\ D62, 084039
  (2000).

\bibitem{GrACE} \url{http://www.caip.rutgers.edu/TASSL/Projects/GrACE/}

\bibitem{GRAstro} \url{http://wugrav.wustl.edu/research/projects/nsgc.html}

\bibitem{Cactus} \url{http://www.cactuscode.org}

\bibitem{Font00} J.~A.~Font, \url{http://www.livingreviews.org/lrr-2003-4}

\bibitem{Gourgoulhon01} E.~Goulgoulhon et al., Phys.\ Rev.\ D63, 064029, (2001)

\bibitem{Taniguchi02} K.~Taniguchi and E.~Gourgoulhon, Phys.\ Rev.\ D66,
  104019, (2002)

\bibitem{Kawamura03} M.~Kawamura et al., astro-ph/0306481.

\bibitem{Shibata03} M.~Shibata et al., Phys.\ Rev.\ D68 (2003) 084020.

\bibitem{Faber03} J.~A.~Faber et al., gr-qc/0312097.

\end{thebibliography}

\end{document}